\documentclass[useAMS,usenatbib,usegraphicx]{mn2e}

\usepackage{epsfig}
\usepackage{amsmath, amssymb,bm}
\usepackage{booktabs}
\usepackage{multirow}
\PassOptionsToPackage{hyphens}{url}\usepackage{hyperref}
\usepackage[hyphenbreaks]{breakurl}

\def \be{\begin{equation}}
\def \ee{\end{equation}}

\def \msun{\rm M_{\odot}}

\def \me{{\dot M_{\rm Edd}}}
\def  \le{{L_{\rm Edd}}}
 
\begin{document}

\topmargin=-0.7in

\title[AGN Discs and the Self--Gravity Limit]{AGN Light Echoes and the Accretion Disc Self--Gravity Limit}

\author[Lobban \& King] 
{\parbox{5in}{Andrew Lobban$^1$ \& Andrew King$^{2,3,4}$
}
\vspace{0.1in} \\ 
$^1$European Space Agency (ESA), European Space Astronomy Centre (ESAC), Camino Bajo del Castillo s/n, \\ 28692 Villanueva de la Cañada, Madrid, Spain\\
$^2$ School of Physics \& Astronomy, University
of Leicester, Leicester LE1 7RH UK\\ 
$^3$ Astronomical Institute Anton Pannekoek, University of Amsterdam, Science Park 904, NL-1098 XH Amsterdam, The Netherlands \\
$^{4}$ Leiden Observatory, Leiden University, Niels Bohrweg 2, NL-2333 CA Leiden, The Netherlands
}

\date{Accepted for publication in MNRAS on 13 January 2022}

\maketitle

\begin{abstract}
Accretion disc theory predicts that an AGN
disc becomes self--gravitating and breaks up into stars at an outer radius 
$R_{\rm sg} \simeq 12$ light-days, 
with effectively no free parameter. We present evidence that the 
longest observed AGN light echoes are all close to 12\,d in the AGN rest frames. 
These observations give a stringent test of AGN disc theory. Further 
monitoring should offer insight into the formation angular momentum of the gas 
forming the disc. For distant AGN, observed lags significantly longer than 12\,d give lower 
limits on their redshifts.
\end{abstract}


\section{Introduction}

Astronomers now generally agree that the centre of almost every galaxy contains
a supermassive black hole (SMBH). How this hole came to grow, radiate as an 
active galactic nucleus (AGN),  and 
influence its host galaxy, are topics of considerable current interest. 

One aspect of this is trying to understand the structures surrounding the hole.
Closest to the hole is the accretion disc, whose gas radiates the 
accretion luminosity as it spirals in. This gas ultimately accretes to the hole, growing 
its mass and modifying its spin. The existence of 
a disc of some kind, where the infalling gas loses angular momentum to gas 
further out, is inevitable -- we know from the Soltan relation \citep{Soltan82} that the hole gained most of its mass through luminous accretion of gas,
which must have occupied a region at least as large as the current hole's sphere of
influence: 
\be 
R_{\rm inf} = \frac{2GM}{\sigma^2}.
\label{inf}
\ee
Here, $M$ is the current hole mass, and $\sigma$ is the central velocity 
dispersion of the host galaxy. For typical $M, \sigma$ we have $R_{\rm inf} 
\gtrsim 1 - 10\,{\rm pc}$, and it is implausible that this gas somehow lost all
angular momentum and fell radially into the hole from large distance. 

As gas accretes inwards in the disc, the outer parts of the disc take up its
angular momentum and move outwards to accumulate at larger radius. But there is a limit to this process,
since the outer parts eventually become self--gravitating at a sufficiently large mass.
The gaseous disc tends to break up into stars at this point, and
no longer spreads to larger significantly larger radii.

The full disc equations \citep{CollinSouffrinDumont90} show that the 
self--gravity radius is:
\be
R_{\rm sg} \simeq 3\times 10^{16}\,{\rm cm},
\label{rsg}
\ee
almost independently of parameters (see the Appendix for a discussion).

Further out from the central SMBH and its accretion disc, it is likely that cooler dusty gas, possibly
gas left over from an earlier accretion event, 
forms a structure with a toroidal topology.
This has inner radius $R_{\rm torus} \sim 1\,{\rm pc}$ and outer radius of order $10$ 
times this.  The largest structure associated with the SMBH is probably the gas 
identified in X-ray absorption spectra as the `warm absorber'. This may have formed
from gas originally surrounding the hole which was swept into a shell by radiation
pressure from the accreting black hole \citep{KingPounds14}, stalling at radii where it
becomes optically thin to electron scattering. It has a
characteristic radius $\sim 1-100\,{\rm pc}$,
rather larger than the black hole's radius of influence, $R_{\rm inf}$. 

The most striking property of any of these suggested structures is that the
disc self--gravity radius has an effectively fixed size for every AGN, virtually independent
of all parameters (cf equation~\ref{rsg}). This unusually precise estimate potentially
offers a powerful observational test of the whole picture of AGN accretion and black hole growth. 

Promisingly, our
own Galaxy does give quantitative supporting evidence. There is at least one 
ring of stars around Sgr A* at a radius slightly
larger than $R_{\rm sg}$ (e.g. \citealt{Genzel03}). This is what we would expect if an earlier accretion 
episode created these stars at its self--gravity limit. As further gas accreted on to
Sgr A*, the stars would have absorbed the angular momentum transported 
outwards to $R_{\rm sg}$ and moved to slightly larger radii. 

It is clear that an extension of this kind of test to a large number of galaxies is 
currently very unlikely. In the next Section we discuss a different test, which can probe the 
apparently universal relation (equation~\ref{rsg}).

\section{Lags}

In recent years there has been considerable progress in using light echoes (reverberation mapping) to investigate AGN accretion-disc structure (see \citealt{CackettBentzKara21} for a review). Reverberation techniques are well established for studying the geometry and kinematics of diffuse emission in the more distant broad-line region (BLR). Here, responses in broad emission lines are measured with respect to continuum variations on time-scales as long as a few to hundreds of days, for example in the case of PG quasars (e.g. \citealt{Peterson04}). We can tentatively understand these very long lags as coming from reprocessing from the AGN BLR region with characteristic scales, $R_{\rm BLR}$.

However, similar techniques are now being routinely applied to mapping the accretion disc itself. Here, the fundamental idea is to analyse correlated variations of the emission signal in distinct continuum bands. The emission observed in different wavelength bands likely predominantly arises from different sites -- e.g. with the X-rays originating close to the central SMBH, while the thermal UV/optical disc emission arises from regions at larger radii. Disc reverberation studies are typically discussed in terms of `accretion disc reprocessing' models. Here, the X-rays -- which are produced via inverse Comptonisation in a corona of hot electrons, close to the SMBH \citep{HaardtMaraschi91} -- may become thermally `reprocessed' via re-irradiation of the accretion disc \citep{GuilbertRees88}. This results in the production of longer-wavelength photons, where the variability is coupled -- e.g. an increase in X-ray emission results in an increase in UV/optical flux with a characteristic time delay, governed by the light-crossing time between the respective emission sites. As the time delay directly relates to the distance between the emission sites, this method can be used to probe the structure of the disc and test various predictive models of accretion.

In a standard, optically--thick accretion disc, the radial temperature dependence is:

\begin{equation} T(r) \approx 6.3 \times 10^{5} \left(\frac{\dot{M}}{\dot{M}_{\rm Edd}}\right)^{1/4} M_{8}^{-1/4} \left(\frac{r}{R_{\rm S}}\right)^{-3/4},  \label{eq:radial-temp} \end{equation}

where $T$ is the temperature in $K$, $(\dot{M}/\dot{M_{\rm Edd}})$ is the accretion rate as a function of the Eddington rate, $M_{8}$ is the mass of the 
central black hole in units of $10^{8}$\,M$_{\odot}$, and $(r/R_{\rm S})$ is the radius from the SMBH as a function of the Schwarzschild radius (see equations 
3.20 and 3.21 of \citealt{Peterson97}). Combining equation~\ref{eq:radial-temp} with Wien's Law ($\lambda_{\rm max} = 2.9 \times 10^{7}/T$, where $\lambda$ is 
the peak effective wavelength of the given filter in units of \AA), provides us with an estimate of the temperature of the disc at a
given radius. For a black hole with a mass of $10^{8}$\,M$_{\odot}$, 
accreting at 
$L/L_{\rm Edd} = 0.1$, this would place the expected emission-weighted radius of the {\it B}-band emission (typically $\sim$4\,400\,\AA) at 
202.9\,$R_{\rm S}$, or $\sim$2.3\,light-days from the central source. Alternatively, one may replace Wien's Law with a more realistic flux-weighted radius (i.e. assuming that $T \propto R^{-3/4}$), which typically predicts lag separations that are up to a 
factor of a few smaller (see \citealt{Edelson19}). In either case, the longer-wavelength emission is expected to be delayed, $\tau$, with respect to the shorter-wavelength emission following the relation: $\tau \propto \lambda^{4/3}$ \citep{CackettHorneWinkler07}.  This relation is expected to hold whether the 
disk is heated internally or externally (also see \citealt{Netzer13}).

The above assumes that the source accretes in the form of a standard, optically-thick $\alpha$-disc (i.e. $H/R = 0.01$, $\alpha = 0.1$, and with a central, compact source of X-rays lying at $6$\,$r_{\rm g}$ above the mid-plane). For the disc reprocessing model to be viable requires that the process is energetically consistent. Thermal emission in the UV/optical bands is typically more luminous than the X-ray emission. As such, the variations in the lower--luminosity X--ray band must be sufficiently large enough in amplitude to drive smaller variations in the higher-luminosity, longer-wavelength bands. In the AGN studied in this context to date, this criterion is usually satisfied. Other caveats include requirements that the observed variability is smooth and fast enough that sufficient maxima and minima are sampled in the light curves to detect the variability on the required time-scales. The variability must also be of high enough amplitude to be detected over the intrinsic host galaxy emission, which is typically expected to be almost constant on these time-scales.

\subsection{The cross-correlation function}

When searching for correlated time delays between two separate time series, we typically would like to use the cross-correlation function [CCF (see \citealt{BoxJenkins76, GaskellPeterson87, MaozNetzer89, KoratkarGaskell91})], where one measures the strength of the correlation coefficient as a function of time delay, $\tau$. By aligning the most prominent maxima and minima, one can maximize the value of the CCF, and arrive at the average time delay. However, in astronomy, data are typically not evenly sampled. This is often the case when performing multiwavelength monitoring campaigns on AGN with either ground--based or space--based observatories, for reasons ranging from weather to scheduling constraints. As such, one workaround is to approximate the CCF using the discrete correlation function (DCF; \citealt{EdelsonKrolik88}), whereby sets of data pairs are binned according to some $\Delta\tau$ (typically set at half of the average observed sampling rate).  We note that when computing the DCF, the number of points falling inside each time bin can vary greatly, depending on the sampling of the two time series. As such, the width of the bins can be varied to achieve statistical significance in each bin. This is known as the $z$-transformed DCF (ZDCF; \citealt{Alexander97}).  Alternatively, one can achieve regular sampling by linearly interpolating between consecutive data points in one time series in order to measure the CCF for any arbitrary value of $\tau$ between the interpolated values in the first series with the real values in the second (and vice-versa). This is known as the interpolated correlation function [ICF (\citealt{GaskellSparke86, WhitePeterson94})].  The advantage of the ICF is that it can achieve greater resolution than its discrete counterpart, assuming that the variability functions in the two time series are relatively smooth (see \citealt{Peterson01}). Meanwhile, uncertainties on the lag measurements can be estimated using the `flux randomisation / random subset selection' (FR/RSS) method outlined in \citet{Peterson98}.  We note that it is common practice in the literature to quote both the peak, $\tau_{\rm peak}$, and the centroid, $\tau_{\rm cent}$, of the CCF, where $\tau_{\rm cent}$ is typically determined from all $r_{\rm CCF}$ values that are $>0.8r_{\rm peak}$.

Other methods are also gaining popularity for searching for time-dependent correlations between continuum bands. For example, \textsc{javelin} \citep{ZuKochanekPeterson11}, which builds upon damped random-walk models for characterising quasar variability \citep{KellyBechtoldSiemiginowska09, Kozlowski10, Macleod10}. The idea of \textsc{javelin} is to first fit a damped random-walk model to the reference time series. This model is then smoothed and shifted to fit the comparison time series, allowing the best-fitting time delay to be obtained. The model is found to obtain robust measurements of inter-band time delays (e.g. \citealt{Jiang17}) and compares well with CCF measurements from disc reverberation studies, although with some discrepancies in the lag uncertainties (e.g. \citealt{Edelson19}). Other models have also recently been developed with the aim of testing disc reverberation mapping based on specific geometries (e.g. the `lamppost model' of the \textsc{cream} code; \citealt{StarkeyHorneVillforth16}).

\subsection{Multiwavelength observations} \label{sec:lags}

A number of intensive long-term multiwavelength monitoring campaigns have now been performed on AGN with coordinated ground-based programs and space-based missions such as the {\it Rossi X-ray Timing Explorer} (RXTE; \citealt{Bradt93}) and the {\it Neil Gehrels Swift Observatory} (hereafter: {\it Swift}; \citealt{Gehrels04}).  While the number of campaigns to date is limited because of the long time-scales needed to sample enough variability, they are growing in number thanks to series of interesting results. In particular, challenges to the disc reprocessing scenario have arisen due a number of measured time delays that appear larger than expected from predictions of standard accretion disc models (e.g. \citealt{CackettHorneWinkler07, Edelson15}). As this suggests that discs appear hotter than expected at a given radius, the implication is that discs may be larger in extent than predicted. Such results have been independently supported by microlensing studies of quasars (e.g. \citealt{Mosquera13}), with some results suggestive of disc sizes that are up to four times larger than predicted \citep{Morgan10}. However, we note that other systematic uncertainties are in play, such as SMBH masses and accretion rates, and so the picture is not entirely clear.

Curiously, it also commonly observed that the optical-UV correlation is smoother and stronger than when compared to the X-rays. This may be expected if the optical- and UV-emitting regions are significantly larger in spatial extent than the compact X-ray source. Additionally, faster modes of variability are often imprinted on X-ray light curves, for example through absorption from ionised material in the line of sight. Typically, this predominantly affects the spectrum at energies $< 2$\,keV and so, in some cases, the problem can be alleviated by considering X-rays at harder energies (e.g. $> 2$\,keV). The less well-constrained variability with the X-rays has also led to suggestions that the X-rays that we see are not the same band that drives the reprocessing mechanism -- i.e. the driving band could be higher in energy than the observable 0.3--10\,keV band of the {\it Swift} X-ray Telescope (see \citealt{Edelson19}) or lower in energy in the unobservable Extreme Ultraviolet (EUV) band (see \citealt{GardnerDone17}).

Nevertheless, since $R_{\rm sg}$ is thought to be an upper limit to the size of 
every AGN accretion disc, and there is a strong selection effect against 
measuring lags from accretion discs highly inclined to the line of sight,
it follows that all reverberation lags from disc gas should obey:
\be
\tau < \frac{R_{\rm sg}}{c} \simeq 10^6\,{\rm s} \simeq 12\,{\rm d} \label{eq:12-day}
\ee
in the AGN rest frame.
Lags from light echoes from surrounding stars may be slightly longer that this.
Lags from the next bigger structure (the torus) should be significantly longer than
this, i.e.:
\be 
\tau \gtrsim \frac{0.1\,{\rm pc}}{c} \gtrsim 120\, {\rm d}. \label{eq:torus_lags} 
\ee 

As such, for any AGN accretion disc for which reprocessing is dominated by the light-travel time, we should not expect to observe rest-frame time delays much greater than $\sim$12\,d from the central driving continuum source, regardless of properties such as SMBH mass. Here, we searched the literature for reported disc-reverberation time delays from known, intensive, multiwavelength monitoring campaigns of AGN.  \citet{KammounPapadakisDovciak21} define a sample of eight such sources with simultaneous X-ray/UV/optical data: Mrk\,142, Mrk\,509, NGC\,2617, NGC\,4151, NGC\,4593, NGC\,5548, NGC\,7469, and Fairall\,9. We summarise the prominent published lags from these campaigns in Table~\ref{tab:results}, where the appropriate references for each campaign are cited in the caption. We also include MCG$+$08$-$11$-$011, although this study did not have simultaneous X-ray data. For NGC\,2617, we point the reader towards the results of \citet{Shappee14}, where robust time lags were obtained, extending up to $\sim$8\,d between the X-ray and the near-infrared (NIR) bands. However we do not include these results as the authors were forced to compute only time delays of $< 10$\,d with \textsc{javelin} to avoid spurious peaks arising from two light-curve peaks separated by $\sim$15\,d.  We note that some authors quote time delays with respect to different reference bands. Where possible, we quote the published values with respect to the X-ray band, and additionally, any other time delays of interest -- i.e. the longest detected lags and/or those between continuum bands which have the largest separation in wavelength. We note that, for the results from \citet{Edelson19}, we quote the results with respect to the hard X-ray band (HX: 1.5-10\,keV), where they attempt to alleviate any effects of soft X-ray absorption. We also note that the majority of these time delays are given in the observed frame -- however, the effect of comsological redshift on the magnitude of time delays obtained from nearby, low-redshift AGN is small.  Nevertheless, we ensure to convert all reported time delays to the appropriate rest frame, if not already done so.

\begin{table*}
\centering
\begin{tabular}{l c c l c c}
\toprule
Source & Bands & $\tau$ (d) & Source & Bands & $\tau$ (d) \\
\midrule
$^{*}$1E\,0754.6$+$3928$^{a}$ & {\it B}\,(4\,353\,\AA)-{\it V}\,(5\,477\,\AA) & $12.0\pm3.6$ & Mrk\,509$^{a,k}$ & {\it HX}-{\it W2}\,(1\,928\,\AA) & $-4.94^{+1.39}_{-2.02}$ \\
& {\it B}\,(4\,353\,\AA)-{\it R}\,(6\,349\,\AA) & $17.9^{+3.2}_{-3.4}$ & & {\it B}\,(4\,353\,\AA)-{\it R1}\,(7\,865\,\AA) & $7.26^{+1.63}_{-1.87}$ \\
& {\it B}\,(4\,353\,\AA)-{\it R1}\,(7\,865\,\AA) & $18.8^{+3.3}_{-4.2}$ & & {\it B}\,(4\,353\,\AA)-{\it I}\,(8\,797\,\AA) & $8.82^{+1.78}_{-2.39}$ \\
& {\it B}\,(4\,353\,\AA)-{\it I}\,(8\,797\,\AA) & $15.9^{+4.2}_{-4.8}$ & Mrk\,817$^{l}$ & {\it W2}\,(1\,928\,\AA)-{\it z}\,(8\,700\,\AA) & $5.15^{+1.14}_{-1.32}$ \\
3C\,390.3$^{a}$ & {\it B}\,(4\,353\,\AA)-{\it R}\,(6\,349\,\AA) & $7.68^{+1.39}_{-1.84}$ & NGC\,2617$^{g}$ & {\it X}-5\,100\,\AA & $2.47^{+0.81}_{-0.88}$ \\
& {\it B}\,(4\,353\,\AA)-{\it R1}\,(7\,865\,\AA) & $8.97^{+1.43}_{-1.89}$ & & 5\,100\,\AA-{\it z}\,(8\,982\,\AA) & $0.86^{+0.64}_{-0.61}$ \\
& {\it B}\,(4\,353\,\AA)-{\it I}\,(8\,797\,\AA) & $8.92^{+1.40}_{-2.62}$ & NGC\,3516$^{d}$ & {\it X}-{\it W2}\,(1\,928\AA) & $1.6 \pm 1.5$ \\
Ark\,120$^{b}$ & {\it X}-{\it V}\,(5\,468\,\AA) & $11.9 \pm 7.3$ & NGC\,3783$^{h}$ & {\it X}-{\it B}\,(4\,500\,\AA) & $6.6^{+7.2}_{-6.0}$ \\
& {\it X}-{\it i}\,(8\,797\,\AA) & $12.3 \pm 4.8$ & NGC\,4051$^{m}$ & {\it X}-{\it R}\,(6\,500\,\AA) & $2.8^{+0.6}_{-0.7}$ \\
\textsuperscript{{\textdagger}}DES\,J0328$-$2738$^{c}$ & {\it g}\,(4\,720\,\AA)-{\it i}\,(7\,835\,\AA) & $7.7^{+1.1}_{-1.2}$ & NGC\,4151$^{k}$ & {\it HX}-{\it W2}\,(1\,928\,\AA) & $3.32^{+0.35}_{-0.27}$ \\
Fairall\,9$^{d,e}$ & {\it X}-{\it V}\,(5\,468\,\AA) & $4.2\pm2.8$ & & {\it W2}\,(1\,928\,\AA)-{\it V}\,(5\,468\,\AA) & $0.96^{+0.51}_{-0.50}$ \\
& {\it W2}\,(1\,928\,\AA)-{\it z}$_{s}$\,(8\,700\,\AA) & $7.03^{+1.19}_{-1.01}$ & NGC\,4593$^{k}$ & {\it HX}-{\it W2}\,(1\,928\,\AA) & $0.60^{+0.11}_{-0.12}$ \\
IRAS\,13224$-$3809$^{d}$ & {\it X}-{\it W2}\,(1\,928\AA) & $6.4 \pm 3.7$ & & {\it W2} (1\,928\,\AA)-{\it V}\,(5\,468\,\AA) & $0.35^{+0.27}_{-0.30}$ \\
MCG$-$06$-$30$-$15$^{f}$ & {\it B}\,(4\,500\,\AA)-{\it J}\,(12\,200\,\AA) & $13.5\pm4.3$ & NGC\,5548$^{a,k}$ & {\it HX}-{\it W2}\,(1\,928\,\AA) & $4.55^{+0.72}_{-1.19}$  \\
MCG$+$08$-$11$-$011$^{a,g}$ & {\it u}\,(3\,449\,\AA)-5\,100\,\AA & $-0.66^{+0.60}_{-0.68}$ & & {\it W2}\,(1\,928\,\AA)-{\it V}\,(5\,468\,\AA) & $1.41^{+0.43}_{-0.41}$ \\
& 5\,100\,\AA-{\it z}\,(8\,982\,\AA) & $1.94^{+0.48}_{-0.57}$ & & {\it B}\,(4\,353\,\AA)-{\it R}\,(6\,349\,\AA) & $10.7^{+1.4}_{-0.8}$ \\
MR\,2251$-$178$^{h}$ & {\it B}\,(4\,500\,\AA)-{\it J}\,(12\,200\,\AA) & $9.0^{+4.0}_{-3.5}$ & & {\it B}\,(4\,353\,\AA)-{\it R1}\,(7\,865\,\AA) & $8.5^{+1.5}_{-1.1}$ \\
Mrk\,110$^{i}$& BAT-{\it g}\,(4\,770\,\AA) & $9.43^{+1.27}_{-1.53}$ & & {\it B}\,(4\,353\,\AA)-{\it I}\,(8\,797\,\AA) & $10.2^{+2.1}_{-1.0}$ \\
& {\it g}\,(4\,770\,\AA)-{\it z}\,(9\,132\,\AA) & $8.01^{+1.10}_{-0.70}$ & NGC\,6814$^{n}$ & {\it X}-{\it B}\,(4\,329\AA) & $2.6^{+1.3}_{-1.5}$ \\
& \**{\it X}-{\it B}\,(4\,329\AA) & $7.75 \pm 3.35$ & NGC\,7469$^{o}$ & {\it X}-6\,962\,\AA & $2.2 \pm 0.8$ \\
Mrk\,142$^{j}$ & {\it X}-{\it W2}\,(1\,928\,\AA) & $2.09^{+0.40}_{-0.34}$ & \textsuperscript{{\textdagger}}SDSS J1415$+$5409$^{p}$ & {\it g}\,(4\,750\,\AA)-{\it i}\,(7\,630\,\AA) & $17.7^{+24.9}_{-9.1}$  \\
& {\it W2}\,(1\,928\,\AA)-{\it z}\,(8\,700\,\AA) & $2.08^{+0.64}_{-0.73}$ & \textsuperscript{{\textdagger}}SDSS J1420$+$5216$^{p}$ & {\it g}\,(4\,750\,\AA)-{\it i}\,(7\,630\,\AA) & $12.8^{+20.5}_{-5.5}$ \\
\midrule
\multicolumn{6}{c}{Pan-STARRS Medium Deep Field \citep{Jiang17}} \\
J022144.75$-$033138.8 & {\it g}-{\it z} & $13.8 \pm 0.3$ & J142336.76$+$523932.8 & {\it g}-{\it z} & $11.7 \pm 0.8$ \\
J084536.18$+$453453.6 & {\it g}-{\it z} & $12.0 \pm 3.3$ & J141104.86$+$520516.8 & {\it g}-{\it z} & $14.2 \pm 0.8$ \\
J083836.14$+$435053.3 & {\it g}-{\it z} & $12.6 \pm 0.8$ & J141856.19$+$535844.9 & {\it g}-{\it z} & $14.3 \pm 0.3$ \\
J100421.01$+$013647.3 & {\it g}-{\it z} & $12.1 \pm 0.8$ & J142106.26$+$534406.9 & {\it g}-{\it z} & $11.9 \pm 0.3$ \\
\bottomrule
\end{tabular}
\caption{Selected time lags with the largest magnitudes and/or arising from some of the most prominent long-term monitoring campaigns reported in the literature, up to September 2021.  For each source, the fundamental criteria for selection are those lags which are either observed to be the longest and/or arise from emission bands that have the largest separation in wavelength.  The lower portion of the table shows the longest time delays from the \citet{Jiang17} study (source names are in SDSS format). Refs: $^{a}$\citet{Sergeev05}; $^{b}$\citet{Lobban20}; $^{c}$\citet{Yu20}; $^{d}$\citet{Buisson17}; $^{e}$\citet{HernandezSantisteban20}; $^{f}$\citet{Lira15}; $^{g}$\citet{Fausnaugh18}; $^{h}$\citet{Lira11}; $^{i}$\citet{Vincentelli21} \**[also see Lobban et al. (in prep.)]; $^{j}$\citet{Cackett20}; $^{k}$\citet{Edelson19}; $^{l}$\citet{Kara21}; $^{m}$\citet{Breedt10}; $^{n}$\citet{Troyer16}; $^{o}$\citet{Pahari20}; $^{p}$\citet{Homayouni19}. Note that BAT in the Mrk\,110 result refers to the Burst Alert Telescope on-board {\it Swift} (15--50\,keV). In the case of the Pan-STARRS survey, we quote the eight longest time delays, which are all observed between the {\it g} (4\,810\,\AA) and {\it z} (8\,660\,\AA) bands. All quoted values are obtained from CCF analyses where possible, although in some cases -- e.g. the \citet{Jiang17} Pan-STARRS results -- the \textsc{javelin} code was used.  All lag values are given in the rest frame.  $^{*}$See the text for additional notes on this particular source (Section~\ref{sec:lags}). \textsuperscript{{\textdagger}}Full source names are provided in \citet{Shen15} and \citet{Yu20}.}
\label{tab:results}
\end{table*}

In Table~\ref{tab:results}, we also include results from a sample of 21 AGN published by \citet{Buisson17}, quoting the time delays where a significant correlation was observed.  In addition, we include results from other known multiwavelength monitoring campaigns, including NGC\,3783 \citep{Arevalo09,Lira11}, NGC\,4051 \citep{Breedt10}, MR\,2251$-$178 \citep{Lira11}, MCG$-$6$-$30$-$15 \citep{Lira15}, Mrk\,817 \citep{Kara21}, and Ark\,120, which may have one of the longest published time delays from a nearby AGN to date \citep{Lobban20}.  With \textit{Swift}, the {\it V} band (5\,468\,\AA) is observed to lag behind the X-rays with $\tau = 11.90 \pm 7.33$\,d, while additional ground-based monitoring with the {\it Skynet Robotic Telescope Network}\footnote{\url{https://skynet.unc.edu/}} revealed a long delay between the X-rays and the {\it i} band (8\,797\,\AA) of $\tau = 12.34 \pm 4.83$\,d. Ark\,120 has a larger black hole mass than most sources targeted by these campaigns ($M_{\rm BH} = 1.5 \pm 0.2 \times 10^{8}$\,$M_{\odot}$; \citealt{Peterson04}).  In larger-mass systems, the spatial separation between regions that emit in two given wavelength bands is expected to be larger -- as such, for a given wavelength gap, the time delay is expected to be longer, up to the $\sim$12-d limit stated in equation~\ref{eq:12-day}.  However, the slower variability of such larger-mass systems also requires that much longer observing campaigns are required to obtain robust measurements, hence the large uncertainties on these values.  These uncertainties may be reduced with longer observing campaigns that sample more peaks and troughs in the light curves with which to anchor the CCF.  On that note, {\it Swift} monitoring campaigns of higher-mass AGN do exist -- e.g. PDS\,456 ($M_{\rm BH} \sim 10^{9}$\,$M_{\odot}$: \citealt{Reeves09, Nardini15}) -- however, the UV / optical variability is too slow and/or weak to be detected (e.g. \citealt{Reeves21}).  Additionally, PDS\,456 likely accretes at close to the Eddington limit, thus lengthening the expected time delays between the emission in two given wavelength bands.  Conversely, we note that in low-mass AGN, such as NGC\,4395 ($M_{\rm BH} = 3.6 \times 10^{5}$\,$M_{\odot}$; \citealt{BentzKatz15}), X-ray-to-UV lags of $\sim$470\,s can be detected within a single {\it XMM-Newton} observation on the order of $\sim$1\,d using the Optical Monitor \citep{McHardy16}.

A large sample of 240 quasars in the Pan-STARRS Medium Deep Fields was also studied by \citet{Jiang17}. Despite the absence of X-ray data, the authors detect lags between the {\it g} (effective wavelength: $\sim$4\,810\,\AA) band and the {\it r} ($\sim$6\,170\,\AA), {\it i} ($\sim$7\,520\,\AA), and {\it z} ($\sim$8\,660\,\AA) bands. A handful of sources stand out due to the magnitude of their measured {\it g}-to-{\it z}-band lags, and we include these in Table~\ref{tab:results}. \citet{Jiang17} also typically find lags that are $\sim$2-3 times larger than predicted from disc equations. An additional lag sample was reported by \citet{Jha21} from the {\it Zwicky Transient Facility} survey, with lag detections between the {\it g}, {\it r}, and {\it i} bands found in 19 different AGN.  The longest lag reported in their study occurs in the source denoted as RMID779 and has a magnitude of $\tau = 9.0^{+0.0}_{-2.8}$\,d between the {\it g} and {\it r} bands.  Again, the authors find evidence for larger-than-predicted disc sizes, this time by an average factor of 3.9 across their whole sample.

In addition, a number of large, ground-based surveys have been performed, focusing primarily on optical and infrared bands (as opposed to X-rays).  \citet{Sergeev05} searched for lags in the {\it V}, {\it B}, {\it R}, {\it R}1, and {\it I} bands in 14 AGN using the {\it Crimean Astrophysical Observatory}.  They found lags in their sample ranging from tenths of days to several days.  The most interesting result in the context of this paper occurs in the source 1E\,0754$+$3928.  Here, \citet{Sergeev05} find lag centroids between the {\it B} band and the {\it V}, {\it R}, {\it R}1, and {\it I} bands of $12.0 \pm 3.6$, $17.9^{+3.2}_{-3.4}$, $18.8^{+3.3}_{-4.2}$, and $15.9^{+4.2}_{-4.8}$\,d, respectively (where we have converted the quoted values into the rest frame).  The middle two measurements are clearly greater than 12\,d within the quoted 1$\sigma$ uncertainties.  However, the authors also present 3$\sigma$ uncertainties — the lags then fall in the ranges: $2.2-23.7$, $7.4-29.7$, $5.8-31.2$, and $0.2-30.8$\,d, respectively, consistent with $< 12$\,d.  Furthermore, there is a significant discrepancy between the peaks of the lags and their associated centroids, likely hinting at complex structure within the CCFs.  For the four bands, the (rest-frame-corrected) reported peaks appear to fall at or around the $\sim$12-d limit we discuss here: {\it V} $= 8.6^{+6.6}_{-6.0}$, {\it R} $= 10.2^{+8.8}_{-1.4}$, {\it R}1 $= 11.3^{+8.5}_{-2.0}$, and {\it I} $= 11.8^{+6.9}_{-4.3}$\,d.  We also note that, in their sample, the light curves of 1E\,0754$+$3928 return significantly lower values of the peak correlation coefficient, which ranges from $0.77-0.90$ across the four bands, compared to values that are typically $> 0.95$ in the case of the remaining 13 sources.  Nevertheless, it appears that 1E\,0754$+$3928 may be a source worthy of further investigation in terms of probing the accretion-disc size-limit.

Other large disc-lag surveys that have been studied include those published by \citet{Mudd18} and \citet{Yu20}, as part of the {\it Dark Energy Survey}, who report lags from samples of 15 and 22 luminous quasars, respectively.  In both cases, the authors report lags in the {\it r}, {\it i}, and {\it z} bands, relative to the {\it g} band, using light curves acquired with the Dark Energy Camera \citep{Flaugher15} on the {\it Victor M. Blanco 4m Telescope}.  In both surveys, all reported lags are typically just a few days in magnitude, with the longest lag occurring in DES\,J0328$-$2817: $\tau = 7.7^{+1.1}_{-1.2}$\,d between the {\it g} and {\it z} bands.  Meanwhile, \citet{Homayouni19} recently reported on the Sloan Digital Sky Survey Reverberation-Mapping (SDSS-RM) project, which is monitoring 849 quasars in a single field.  They find 95 quasars with well-defined lags, 33 of which exhibit lags which are detected at the $> 2\sigma$ level, using both CCF methods and \textsc{javelin}.  Aside from 4 spuriously large CCF lags that have very low / negative signal-to-noise and/or are not realised in the complementary \textsc{javelin} methods, \citet{Homayouni19} find lags all typically in the range of a few tenths of days to a few days with the longest reliable lags in their sample having magnitudes that are consistent with $\tau = 12$\,d [i.e. $12.8^{+20.5}_{-5.5}$ and $17.7^{+24.9}_{-9.1}$\,d ($7.8^{+0.6}_{-1.1}$\,d with \textsc{javelin})].

Some authors have since suggested that the observed lags in the UV and optical bandpasses do not, in fact, solely originate in the disc, but instead may arise from reprocessing of EUV emission via bound-free transitions in optically-thick BLR clouds (e.g. see \citealt{KoristaGoad01, GardnerDone17}). Excesses in the 3\,000-4\,000\,\AA\ band of lag spectra have given support to this hypothesis (e.g. \citealt{Cackett18, Edelson15, Edelson19}), where these enhanced time delays at or around the Balmer jump are discussed in the context of additional lag contributions from the BLR. We also note that longer-time-scale variations, separate from disc reprocessing, can also distort CCF measurements (e.g. \citealt{Welsh99}). See \citet{McHardy18} and \citet{Pahari20} for a discussion on filtering out long-time-scale variations in order to separate out the shorter-time-scale lags, which may be more suitable for detecting the effects of disc reprocessing, and also seed photon variations (which may produce opposite-direction lags to disc reprocessing).  However, we note that -- in some cases -- such an approach may filter out the potential to observe any time lags at or around the $\sim$12-d limit.

\begin{figure}
\begin{center}
\rotatebox{0}{\includegraphics[width=8.4cm]{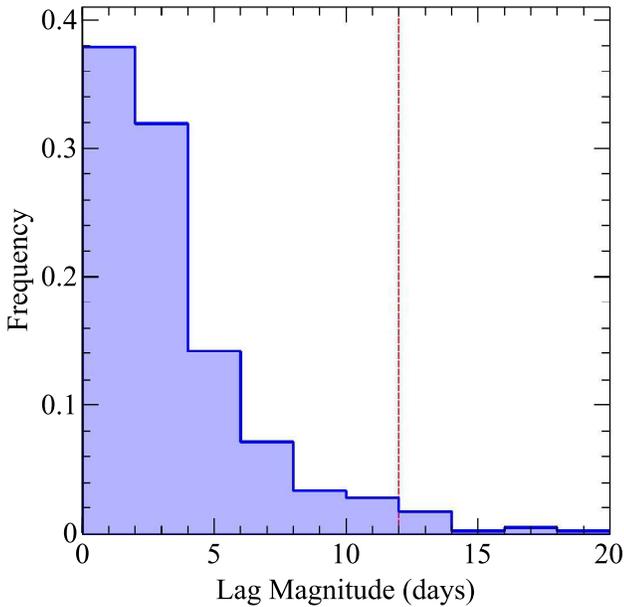}}
\end{center}
\vspace{-15pt}
\caption{Histogram showing observed multiwavelength time lags obtained from the literature.  The bin-size is 2\,d.  The red, vertical dashed line represents the $\sim$12-d cut-off for accretion-disc gas.  There is a clear drop-off at around this value. Note that all lags are plotted in the rest frame, adjusted by the factor $(1+z)$, where $z$ is the redshift. See Section~\ref{sec:lags} for details.}
\label{fig:histogram}
\end{figure}

In Fig.~\ref{fig:histogram}, we generate a histogram of -- to the best of our knowledge -- all continuum-band time delays from intensive multiwavelength monitoring campaigns published in the literature to date.  We include lag measurements from all sources listed in Table~\ref{tab:results}, plus those from surveys such as the 39 Pan-STARRS quasar sub-sample \citep{Jiang17} and those from \citet{Sergeev05}, \citet{Mudd18}, \citet{Homayouni19} and \citet{Yu20}.  The majority of measurements are lag centroids derived from cross-correlation functions, although the Pan-STARRS lags are obtained using \textsc{javelin} due to the highly irregular sampling in the light curves.  We consider all claimed time lags derived from campaigns that have baselines long enough to, in principle, observe time delays of $\sim$12\,d or longer.

We include all claimed lags between continuum bands with a measured time delay that is not consistent with zero.  A handful of measurements are negative relative to the usual convention in the sense that the shorter-wavelength emission lags behind the longer-wavelength emission.  While such measurements may be at odds with the standard disc reprocessing model (see above), we include the magnitude of these lags as, if they do arise from disc gas and are governed by the light-crossing time, they should still be expected to obey $\tau \lesssim 12$\,d. Nevertheless, excluding these few results does not have any significant impact.  For each lag, we also correct for time dilation due to the cosmological redshift, if it is not otherwise stated by the authors that such a correction has already been performed.  Overall, this results in 399 individual rest-frame lag measurements from 123 different AGN / quasars, across a multitude of wavelength bands, all obtained from data acquired by a wide array of space-based and ground-based observatories.

From Fig.~\ref{fig:histogram}, it is clear that the majority of multiwavelength continuum-band time delays published in the literature to date are of the order of a few days or less.  We note that there is a strong bias towards shorter time delays simply because such lags arise from sources with higher variability amplitude on short time-scales.  In other words, most monitoring campaigns have targeted smaller-mass AGN, which vary faster, thus i) making the variability signal easier to pick out, and b) requiring light-curve baselines that are of an achievable length with the current generation of observatories and archival data.  Additionally, the shape of the histogram will be skewed heavily by sources which have been observed more often and with a greater number of wavelength bands, thus revealing more lag detections.  However, the crucial point is that there is currently a clear drop-off as we approach the approximate disc-size limit defined in equation~\ref{eq:12-day} (as noted by the vertical dashed line).  As such, Fig.~\ref{fig:histogram} serves to demonstrate that time delays between accretion-disc continuum bands of around 12\,d or more are either absent or rare, with just a small handful at or around that limit.  Those lags which are observed to be at or slightly longer than 12\,d have either: i) large errors, or ii) smaller errors but have been obtained using \textsc{javelin}, which \citet{Edelson19} state systematically returns smaller errors than CCF method (and from more-irregularly-sampled light curves).

\section{Discussion}

Table~\ref{tab:results} and Fig.~\ref{fig:histogram}
offer remarkable support for the idea that the longest observed X-ray/UV/optical continuum lags in AGN correspond to reverberation from structures very close to the expected disc 
self--gravity 
radius, $R_{\rm sg}$, about 12\,light-days from the central black hole. This value has effectively no 
free parameter,
and is quite distinct in scale from both the AGN torus ($ \gtrsim 100$~d) and the 
still more distant warm absorber gas, making this observation 
a stringent test of theoretical understanding of how AGN accrete. 

Gas accumulates at this radius as successive AGN feeding episodes rid themselves 
of angular momentum. Much of this gas must form stars, and as successive disc 
feeding events 
are probably randomly oriented, the stars must have a roughly spherical
distribution. They move slowly outwards as those with orbits close to the plane 
of the current disc event absorb angular momentum from it, and all the stars 
presumably scatter each other gravitationally. Although the lags are probably mainly 
from the outer accretion disc, there may be a component from this stellar 
distribution. Then, detailed monitoring of lags might potentially constrain the 
initial angular momentum of the disc-feeding events themselves. The orbital period
of these stars is:
\be
P \gtrsim 95 M_{8}^{-1/2}\, {\rm yr},
\ee
where $M_{8}$ is the black-hole mass in units of 
$10^{8}$\,$\msun$. Evidently, detecting the imprint of this is unlikely except 
conceivably for the most massive known AGN. 

The fixed maximum rest-frame lag, $R_{\rm sg}/c$, constitutes a standard
interval. For distant AGN its observed value should be
$12(1+z)$\,d, in principle offering an independent lower limit on the 
redshift, $z$, if the observed lag significantly exceeds 12\,d. 

In equation~(\ref{eq:torus_lags}), we suggest that the next-longest continuum lags should arise from the more distant, dusty toroidal structure, typically at distances of tens to hundreds of light-days from the central black hole (see \citealt{Landt11, Landt19} and references therein), following a well-defined radius-luminosity relation expected for dust sublimation \citep{OknyanskijHorne01}. Curiously, results from a handful of studies appear to be consistent with this picture. For example, \citet{Breedt10} detect secondary peaks between the X-ray and optical bandpasses in NGC 4051. While the primary peak shows the optical emission lagging behind the X-rays with a delay of $\tau = 2.4^{+0.9}_{-1.6}$\,d, the secondary peak is much larger, with $\tau = 38.9^{+2.7}_{-8.4}$\,d.  \citet{Breedt10} find that this secondary time delay is consistent with the dust sublimation radius in NGC\,4051 and so may be the optical signature of the next-largest structure; i.e. the inner edge of the torus.

Likewise, in NGC\,3783, \citet{Lira11} find time delays between the X-ray and {\it B} bands of $\tau = 6.6^{+7.2}_{-6.0}$\,d, while the delays to the NIR bands ({\it J}, {\it H}, and {\it K}) extend to 40-80\,d.  Meanwhile, in the case of MCG$-$6$-$30$-$15, \citet{Lira15} find lags between the optical {\it B} band and the NIR {\it J}, {\it H}, and {\it K} bands of $13.5\pm4.3$, $20.0\pm4.0$, and $26.1\pm3.7$\,d, respectively.  The black-hole mass in MCG$-$6$-$30$-$15 is low ($M_{\rm BH} = 3 \times 10^{6}$\,M$_{\odot}$; \citealt{McHardy05}) -- as such, the accretion-disc lags (e.g. X-ray--{\it B}) are much smaller, and consistent with zero.  Therefore, these are examples of sources in which a significant gap appears between what are likely to be continuum-band lags associated with the accretion disc and the larger-scale dusty torus.  In addition, \citet{Suganuma06} search for time delays between the optical and NIR ({\it J}, {\it H}, {\it K}) bands in three additional sources, also finding much longer time delays: NGC\,5548: 47-53\,d; NGC\,3227: $\sim$20\,d; NGC\,7469: 65-87\,d.  Meanwhile, further dedicated dust-reverberation campaigns and imaging surveys (e.g. \citealt{Koshida14, LyuRiekeSmith19, Minezaki19, Yang20}) continue to build up a large sample of long, correlated time delays from the distant, toroidal structures that surround AGN (also see \citealt{Netzer15}).  Such results help to build up evidence for a picture in which accretion-disc gas may extend out to $\sim$12\,light-days, while continuum-band lags may be observed with significantly longer time delays, relative to the black-hole mass of the source.

In addition, as mentioned in Section~\ref{sec:lags}, there are suggestions in the literature that some observed optical and UV lags may be contaminated by the diffuse continuum arising from optically-thick clouds in the BLR (e.g. \citealt{KoristaGoad01, GardnerDone17}).  In particular, this can affect the 3\,000-4\,000\,\AA\ band (near to the Balmer jump), with excesses in the observed lag spectra often observed here (e.g. \citealt{Cackett18, Edelson15, Edelson19}).  We note that enhanced time delays arising from BLR contributions -- as opposed to continuum reprocessing in the accretion disc itself -- may also imply that, in some cases, true accretion disc lags are even shorter than observed lag measurements imply, lending additional support to the hypothesis of this paper.

Finally, while disc lags that are $\sim$12\,d or longer are very rare in the literature, we note the caveat that, given the nature of most monitoring campaigns, there are many cases where we might not expect to observe such a lag, even if it was present.  For example, most multiwavelength observing campaigns target AGN with lower-mass black holes where the characteristic variability time scales are faster and, therefore, time delays can be detected from shorter observation baselines, placing less of a burden on observatories.  In smaller-mass systems, the distance between respective emission sites is smaller, and so the emission from the outer disc can lie out of the reach of observatories such as {\it Swift}, which can only observe out to the {\it V} band.  Meanwhile, larger-mass systems -- where the outer-disc emission falls within the observable bandpass -- often vary too slowly for significant lags to be detected given a typical campaign length, and so rarely feature in targeted campaigns.  Additionally, the predicted time delays are also a function of accretion rate, with lower-rate sources expected to have a steeper radial-temperature dependence for a given black-hole mass (assuming a standard thin disc), and therefore shorter time delays between two given emission bands.  As such, we note that, while disc lags of 12\,d or more are absent or rare, this is also biased by observers' choices of targets and observing-campaign designs.  However, with some careful choices of source and observing strategies, it should be possible to further robustly test the expected $\sim$12\,light-day limit for accretion-disc gas from self-gravitational considerations.  

\section*{Acknowledgments}

This research has made use of the NASA Astronomical Data System (ADS) and the NASA Extragalactic Database (NED). AL is a current ESA research fellow.


\section*{Appendix: Disc Self--Gravity}\nonumber            

The `vertical' ($z$) component of gravitational force 
from the black hole at the disc surface is $\sim GMH/R^3$ at disc radius $R$, 
where $H$ is the scaleheight, while that from self--gravity is $\sim G\rho H^3/
H^2 \sim G\rho H$, where $\rho$ is the local disc density. Self--gravity begins to
dominate when 
\be 
\rho \sim \frac{M}{R^3}
\label{sg1}
\ee
[\citealt{FrankKingRaine02}, eqn (5.52)]. 
Since the outer parts of the disc carry most of its mass,
we can rewrite this as:
\be 
M_{\rm disc} \sim \pi R^2 H\rho \sim \frac{H}{R}M.
\ee

The full disc equations \citep{CollinSouffrinDumont90} show that in the outer parts of an AGN disc,
\be
\frac{H}{R} \simeq 10^{-3},
\ee
virtually independent of all parameters. This arises because the disc equations imply $H/R = c_s/v_K \propto T_c^{1/2}/v_K \propto
T^{1/2}\tau^{1/2}/v_K \propto (\tau/R)^{1/8} \sim {\rm constant}$,
where $c_s, v_K, T_c, \tau$ are respectively the sound speed, Kepler velocity,
central temperature and vertical optical depth of the disc.
Then, the maximum disc mass is $\sim 10^{-3}M$.

Using the full disc equations gives \citep{CollinSouffrinDumont90}:
\be 
R_{\rm sg} = 3\times 10^{16}\alpha_{0.1}^{14/27}\dot m^{8/27}M_8^{1/27}\,{\rm cm},
\label{rsg2}
\ee
where $\alpha_{0.1} = \alpha/0.1$ and $\dot m = \dot M/\me$ with 
$\me$ the accretion rate giving the Eddington luminosity for black--hole
accretion efficiency, $\eta = 0.1$.

This is another remarkably constant quantity, which arises since the full disc equations
show that $R_{\rm sg}$ is mainly sensitive to $(H/R)^2$, which as we have seen above is, itself, almost constant. 

\section*{Data Availability}

No new data were generated or analysed in support of this research.  The data underlying this article were derived from sources in the public domain; i.e. published literature accessible via the NASA Astrophysics Data System (\url{https://ui.adsabs.harvard.edu/}) and arXiv (\url{https://arxiv.org/archive/astro-ph}).

{}

\end{document}